\documentclass[doublecol]{epl2} 
\usepackage{graphicx}
\usepackage{mathrsfs}
\usepackage[T1]{fontenc} 
\usepackage{bbold}
\usepackage{url}
\usepackage{hyperref}
\usepackage[english]{babel}
\usepackage[utf8]{inputenc}
\usepackage{amsmath}
\usepackage{color}
\usepackage{amsfonts}
\usepackage{amssymb}
\usepackage{mathtools}
\usepackage{placeins}
\usepackage{float}
\usepackage{tabularx}
\usepackage{adjustbox}
\usepackage{caption}
\usepackage{subcaption}

\title{\boldmath Single field slow-roll effective potential from K\"{a}hler moduli stabilizations in type IIB/F-theory}
\shorttitle{\boldmath  Single field slow-roll effective potential from K\"{a}hler moduli stabilizations in type IIB/F-theory} 

\author{Abhijit Let\inst{} \and Arunoday Sarkar\inst{} \and Chitrak Sarkar\inst{} \and Buddhadeb Ghosh \inst{}}
\institute{                    
  \inst{} Centre of Advanced Studies, Department of Physics, The University of Burdwan,\\Burdwan 713 104, India
  
}

\pacs{11.25.Wx}{String and brane phenomenology}
\pacs{98.80.Qc}{Quantum cosmology}
\pacs{98.80.Cq}{Particle-theory and field-theory models of the early Universe (including cosmic pancakes, cosmic strings, chaotic phenomena, inflationary universe, etc.)}

\abstract{ We derive a single field slow-roll inflaton potential in three intersecting $D7$ branes configuration under type IIB/F-theory compactification. Among three resulting K\"{a}hler moduli corresponding to three orthogonal directions, two are stabilized via perturbative corrections in K\"{a}hler potential arising from large volume scenario  ($\alpha'^3$) and four graviton scattering amplitude upto one loop level and the remaining K\"{a}hler modulus is stabilized by KKLT-type non-perturbative correction in superpotential. The symmetric combination of two canonically normalized and perturbatively stabilized K\"{a}hler moduli gives the inflaton field and the anti-symmetric combination manifests as an auxiliary field.  }

\begin{document}

\maketitle
\flushbottom

\begin{figure}[H]\small
    \centering
    \onefigure[width=0.6\linewidth]{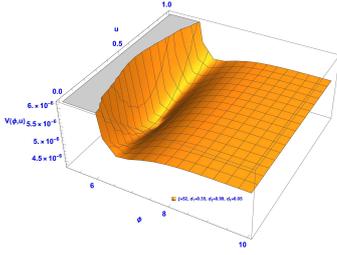}
    \caption{Three dimensional plot of the 2-field inflaton potential of Eq. (\ref{eq:2_field_inf_pot}) against the $\phi$ and $u$. The auxiliary field $u$ remains at its minimum and its component corresponding to the potential has almost no effect, during inflation.}
    \label{fig:3D}
\end{figure}

In recent years, major efforts have been made \cite{Antoniadis:2018hqy,Antoniadis:2018ngr,Antoniadis:2019doc,Antoniadis:2020stf,Basiouris:2020jgp,Basiouris:2021sdf} to obtain cosmological inflation and inflaton potential(s) via K\"{a}hler moduli stabilization in type-IIB/F-theory since the inflationary picture in large scale limit is experimentally connected to the CMBR anisotropy and polarisation data \cite{Planck:2018jri,Planck:2018vyg,BICEP:2021xfz}. In fact, Planck-2018 \cite{Planck:2018jri} has confirmed the efficacy of the single field slow-roll type potentials e.g. the $\alpha$-attractors in explaining the observational bounds with significant precision. Quite obviously, moduli stabilization is essential in order to connect the string compactification to low energy effective theories, such as inflation. Therefore, obtaining a plateau-type inflaton potential via moduli stabilization is an important task in the interface of string theory and cosmology. We have already shown in an earlier publication \cite{Sarkar:2021ird} that the mode-dependent behaviours of cosmological parameters of $\alpha$-attractor potentials conform to the Planck-2018 data to a great extent. This class of potentials originates from the geometry of K\"{a}hler manifold in $\mathcal{N}=1$  minimal supergravity. Motivated by this success, we felt it is pertinent to extract such type of experimentally favoured potential from more fundamental theory viz., the superstring theory through stabilizing the K\"{a}hler moduli fields by quantum corrections in the topology of the internal compact manifold which is a Calabi-Yau threefold.
In the string frame, one is concerned with two potentials: i) the K\"{a}hler potential \cite{Becker:2002nn} $\mathcal{K}$ which generates the metric of the moduli space of the internal manifold, ii) the superpotential \cite{Gukov:1999ya} $\mathcal{W}$  which is generated by world volume fluxes of branes. Usually, $\mathcal{K}$ and $\mathcal{W}$  get contributions from perturbative \cite{Basiouris:2020jgp,Becker:2002nn,Conlon:2005ki, Balasubramanian:2005zx,Antoniadis:2018hqy, Antoniadis:2019rkh,Bobkov:2004cy} and non-perturbative \cite{Basiouris:2020jgp,Witten:1996bn,Kachru:2003aw,Haack:2006cy,Baume:2019sry} effects, respectively. In the process of compactification in string theory, many moduli fields (massless scalars in four dimension) appear which are related to $\mathcal{K}$ and $\mathcal{W}$. The number of moduli fields may be reduced by fluxes \cite{Giddings:2001yu,Grana:2005jc}, $D$ brane compactification \cite{Blumenhagen:2006ci} and orientifold projection \cite{Kachru:2002he}. A single-field inflation is driven by a potential $V(\phi)$, where $\phi$ is the inflaton field. In order to connect string theory to inflation one has to i) derive $V(\phi)$ from a potential $V(\mathcal{K},\mathcal{W})$, which is called the $F$-term potential, and ii) make a transition from $AdS$ space to $dS$ space, the latter having a positive cosmological constant, which is required for inflationary expansion of the universe. Moreover, the potential $V(\phi)$ has to be a slow-roll one. The main motivation of the present work is to obtain a slow-roll potential, $V(\phi)$, incorporating the perturbative ($\alpha'^3$, four-graviton scattering upto genus one and 3 intersecting D7 branes wrapping over 4-cycles), non-perturbative (one instanton/gaugino condensation) corrections in $V(\mathcal{K},\mathcal{W})$. Obtaining a stable $dS$ vacuum from superstring theory is a  challenging task because of the recently proposed swampland conjecture \cite{Ooguri:2006in,Agrawal:2018own} in the context of quantum gravity. But still, efforts have been made to find an effective potential from the stringy perspective, which includes both perturbative\cite{Becker:2002nn,Balasubramanian:2005zx,Antoniadis:2019rkh} and non-perturbative\cite{Kachru:2003aw,Haack:2006cy,Baume:2019sry} elements in the topology of compactified extra dimensions of space-time. Another aspect of this scenario is to uplift the single field effective potential from the $AdS$ to the $dS$ space maintaining a slow-roll plateau. In this paper we have proposed a scheme for deriving a slow-roll $dS$-potential for the inflaton field by stabilizing all the K\"{a}hler moduli and suitably uplifting the $AdS$ minimum to the $dS$ one. Our calculational framework is based on the type IIB superstring theory compactified on a 6d $T^6 /\mathbb{Z}_N$ orbifold limit of Calabi-Yau 3-fold (CY$_3$)\footnote{or an elliptically fibered Calabi-Yau 4-fold in $F$-theory \cite{Iizuka:2004ct}}, which will be designated as $\mathcal{X}_6$ such that the target space is $\mathcal{M}_4\times\mathcal{X}_6$, where $\mathcal{M}_4$ is the 4d Minkowski space. The internal space $\mathcal{X}_6$ is equipped with some non-perturbative objects like $D7$ branes and $O7$ planes. These $O$-planes are necessary to project out the 12d $F$-theory in 10d type IIB theory \cite{Sen:1997gv,Blumenhagen:2010at}. Furthermore, the simplest configuration of three magnetised non-interacting and intersecting $D7$ branes is considered which wrap around the 4-cycles and warp the metric topology of $\mathcal{X}_6$ such that $h^{1,1}=3$. In this set-up, the complexified K\"{a}hler moduli which control the volume of $\mathcal{X}_6$ are expressed as,
\begin{equation}\small
    \rho_k = b_k + i\tau_k, \quad k = 1,2,3.
    \label{eq:kh_moduli}
\end{equation}
$b_k$ is connected with RR $C_4$ potential and $\tau_k$ is identified as 4-cycle volume transverse to the 7-branes. The internal volume of $\mathcal{X}_6$ can be expressed in terms of $\tau_{1,2,3}$ as,
\begin{equation}\small
    \mathcal{V}=\sqrt{\tau_1\tau_2\tau_3}=\sqrt{\frac{(\rho_1 - \bar{\rho_1})(\rho_2 - \bar{\rho_2})(\rho_3 - \bar{\rho_3})}{(2i)^3}},
    \label{eq:volume}
\end{equation} where we have assumed that the intersection number of the branes is one. Branes, wrapping the $\mathcal{X}_6$, produce generalised fluxes \cite{Blumenhagen:2006ci,Giddings:2001yu} threading the 4-cycles of the internal manifold arising due to some potentials $C_p$ and the associated form fields $F_p = dC_{p-1}$, $p=0,2,4$, complexified axion-dilaton $S=C_0 + ie^{-\varphi}$, where $\varphi$ is the dilaton field which is related to the string coupling constant as $\langle\varphi\rangle=\ln{g_s}$; $F_3 = dC_2$, the Kalb-Ramond field $B_2$, $H_3=dB_2$ and $G_3=F_3 - SH_3$. These fields depend on $h^{2,1}$ complex structure moduli $z_a$ which dictate the shape of $\mathcal{X}_6$. The CY$_3$ being a compact K\"{a}hler manifold (which means it is an orientable Riemann surface having finite volume) with vanishing first Chern class and Ricci flatness \cite{Candelas:1985en} admits a non-zero closed holomorphic (3,0) form $\Omega (z_a)$ \cite{Candelas:1990pi} (i.e. $d\Omega (z_a)=0$) everywhere, which is a non-trivial element of Hodge-de Rham cohomology group $H^{3,0}$. The 3-form field $G_3$ and $\Omega (z_a)$ together define the compatibility of CY$_3$ with supersymmetry \cite{Candelas:1985en,Giddings:2001yu} through a flux-generated tree level superpotential \cite{Gukov:1999ya},
\begin{equation}\small
    \mathcal{W}_0 (S,z_a )=\int_{\mathcal{X}_6}G_3(S,z_a) \wedge \Omega (z_a),
\end{equation} which is a holomorphic function of $z_a$ and $S$.
Such type of potential is also described in the $\mathcal{N}=1$ supergravity \cite{Freedman:2012zz} except that, in that case it will be a function of superfields. The supersymmetric constraints require that $z_a$ and $S$ should be supersymmetrically stabilized acquiring large masses at supersymmetric minimum. Therefore, the covariant derivative of $\mathcal{W}_0$ w.r.t. $z_a$ and $S$ must vanish, i.e.
\begin{equation}\small
    \mathcal{D}_S \mathcal{W}_0 = \mathcal{D}_{z_a} \mathcal{W}_0 = 0
    \label{eq:SUSY_stab_ax_d}
\end{equation}
where, $\mathcal{D}\mathcal{W}_0=\partial\mathcal{W}_0 + \mathcal{W}_0 \partial \mathcal{K}$. $\partial\mathcal{K}$ is the connection on the moduli space of CY$_3$. This stabilization ensures that those moduli fields can never appear in large four dimensions. Now, so far as the K\"{a}hler structure (which is the complex structure with a Riemannian metric) of $\mathcal{X}_6$ is concerned, the metric of the moduli space $\mathcal{M} (\mathcal{X}_6) =\mathcal{M}^{2,1} (\mathcal{X}_6)\times \mathcal{M}^{1,1} (\mathcal{X}_6) $, $\mathcal{K}_{I\bar{J}}$ being an exact 2-form i.e. it is derivable (at least locally) from a scalar potential called K\"{a}hler potential $\mathcal{K}_0$,
\begin{equation}\small
    K_{I\bar{J}}=\partial_I \partial_{\bar{J}}\mathcal{K}_0.
\end{equation}
Here $\mathcal{K}_0$ is the classical version of the K\"{a}hler potential and it depends on three moduli viz., $S$, $z_a$ and $\rho_k$ as \cite{Candelas:1990pi,Giddings:2001yu},
\begin{equation}\small
\begin{split}
    \mathcal{K}_0 (S,z_a,\rho_k)=&-\sum_{k=1}^{3} \ln(-i(\rho_k - \bar{\rho}_k))-\ln (-i(S-\bar{S}))\\&-\ln(-i\int \Omega \wedge \bar{\Omega}).
\end{split}
\end{equation}
Using Eq. (\ref{eq:volume}) we get,
\begin{equation}\small
    \mathcal{K}_0 (S,z_a,\rho_k) = -2\ln\mathcal{V}-\mathcal{K}_0 (z_a,S),
    \label{eq:tree_kahler}
\end{equation} where a constant factor $\ln 8$ has been absorbed in $\mathcal{K}_0 (z_a, S)$. $\mathcal{K}_0(S,z_a,\rho_k)$ satisfies an interesting condition,
\begin{equation}\small
    \sum_{k,k' \in  \mathcal{M}^{1,1}({\mathcal{X}_6}) }^{h^{1,1} = 3} {\mathcal{K}_0}^{{\rho_k}\bar{\rho}_{k'}} \partial_{\rho_k} \mathcal{K}_0 \partial_{\bar{\rho}_{k'}} \mathcal{K}_0 = 3
    \label{eq:no_scale}
\end{equation} called the `no-scale' structure, which is necessary to maintain the supersymmetry \cite{Freedman:2012zz}. The superpotential $\mathcal{W}_0$ and the K\"{a}hler potential $\mathcal{K}_0(S,z_a,\rho_k)$ together provide a 4d effective potential called the $F$-term potential,
\begin{equation}\small
\begin{split}
    V_F &= e^{\mathcal{K}_0}\sum_{I,J \in \mathcal{M}({\mathcal{X}_6})}\left( {\mathcal{K}_0}^{I\bar{J}} \mathcal{D}_I \mathcal{W}_0 \mathcal{D}_{\bar{J}} \mathcal{\bar{W}}_0 - 3\mathcal{W}_0 \mathcal{\bar{W}}_0\right)\\
    &=e^{\mathcal{K}_0}{\mathcal{K}_0}^{S\bar{S}}\mathcal{D}_S \mathcal{W}_0 \mathcal{D}_{\bar{S}}\bar{\mathcal{W}_0}\\&+e^{\mathcal{K}_0}\sum_{a,b \in \mathcal{M}^{2,1}({\mathcal{X}_6})}^{h^{2,1}} {\mathcal{K}_0}^{z_a \bar{z}_b} \mathcal{D}_{z_a} \mathcal{W}_0 \mathcal{D}_{\bar{z}_b}\bar{\mathcal{W}_0} \\
    &+e^{\mathcal{K}_0}(\sum_{k,k' \in \mathcal{M}^{1,1}({\mathcal{X}_6})}^{h^{1,1}=3} {\mathcal{K}_0}^{\rho_k \bar{\rho}_{k'}} \mathcal{D}_{\rho_k} \mathcal{W}_0 \mathcal{D}_{\bar{\rho}_{k'}}\bar{\mathcal{W}_0}\\& - 3\mathcal{W}_0 \bar{\mathcal{W}_0})\\
\end{split}
\label{eq:no_scale_v_F}
\end{equation}
which vanishes at classical level due to Eqs. (\ref{eq:SUSY_stab_ax_d}) and (\ref{eq:no_scale}). Therefore, the K\"{a}hler moduli are not fixed at tree level leading to the `moduli stabilization problem'. In order to avoid this problem we have to come out of the classical description and turn on quantum corrections to break the supersymmetric no-scale structure of the K\"{a}hler potential which will give a non-zero $F$-term potential. Let us first focus on the non-perturbative contributions of the K\"{a}hler moduli to $\mathcal{W}_0$ as \cite{Witten:1996bn}
\begin{equation}\small
    \mathcal{W}(S,z_a,\rho_k)=\mathcal{W}_0 (S,z_a) + \sum_{k=1}^3 A_k(z_a) e^{i\alpha_k \rho_k}
    \label{eq:corrected_W}
\end{equation}
arising from various non-perturbative effects like gaugino condensation \cite{Haack:2006cy} and instanton correction \cite{Baume:2019sry} where $\alpha_k$'s are small constants. According to the large volume scenario (LVS) \cite{Balasubramanian:2005zx}, in non-compact limit, all 4-cycles do not expand to infinity, rather, at least one of the 4-cycles must be smaller than others. Also Ref. \cite{Bianchi:2011qh} says that due to certain choices of world volume fluxes in the presence of $E3$ magnetised branes, some of the K\"{a}hler moduli will have non-vanishing contributions to superpotential, which help in applying perturbative string loop effects in K\"{a}hler potential as explained in \cite{Antoniadis:2019rkh}. In our framework we consider $\tau_1$ to be smallest among $\tau_{1,2,3}$ and thus suppressing the effects of larger $\rho$'s we get from Eq. (\ref{eq:corrected_W})
\begin{equation}\small
\begin{split}
    \mathcal{W}(S,z_a,\rho_k)&=\mathcal{W}_0 (S,z_a)+A(z_a)e^{i\alpha \rho_1}\\&=\mathcal{W}_0 (S,z_a)+A'(z_a)e^{-\alpha \tau_1},
\end{split}
    \label{eq:cor_super_pot}
\end{equation} where $A'(z_a)=A(z_a)e^{i\alpha b_1}$. The non-renormalization theorem \cite{Burgess:2005jx} forbids us to modify the superpotential by perturbative corrections. The tree level K\"{a}hler potential is perturbatively corrected through the stabilization of the volume term in Eq. (\ref{eq:tree_kahler}) by the classical $\alpha'^3$ correction \cite{Becker:2002nn} and quantum string loop effects due to multi-graviton scattering \cite{Antoniadis:2019rkh}. The second type of correction is related to the 4d localized Einstein-Hilbert term (second part of Eq. (\ref{eq:grav_action})), originating through the process of compactification from the 10d effective action in gravitational sector given in Ref. \cite{Antoniadis:2019rkh}, containing higher derivative objects\footnote{Higher derivative terms like $\nabla ^4 R^4$ are neglected as their effects are very small in four-graviton scattering amplitude \cite{Green:2008uj,Basu:2007ru}.} like $R^4$ which is also proportional to the non-zero Euler number ($\chi$) of the internal manifold. In this way the 10d effective action reduces to (see \cite{Antoniadis:2019rkh,Basiouris:2020jgp} for details),
\begin{equation}\small
\begin{split}
    S_{\mathrm{grav}}&=\frac{1}{(2\pi)^7 \alpha'^4}\int_{\mathcal{M}_4\times\mathcal{X}_6} e^{-2\varphi} R_{(10)}\\&+ (\alpha')^3\frac{\chi}{(2\pi \alpha')^4}\int_{\mathcal{M}_4}(2\zeta (3)e^{-2\varphi} + 4\zeta (2))R_{(4)},
\end{split}
\label{eq:grav_action}
\end{equation}
where $R_{(10)}=R\wedge e^8$ and $R_{(4)}=R\wedge e^2$. Here $R$ is the Ricci curvature 2-form with $e$ being some generalized basis vector over $\mathcal{X}_6$ (also called \textit{vielbein}),  and
\begin{equation}\small
    \chi = \frac{3!}{(2\pi)^3}\int_{\mathcal{X}_6}R\wedge R\wedge R .
\end{equation}
The terms associated with $\zeta (3)$ and $\zeta (2)$ come from genus zero scattering amplitude which is actually analogous to $\alpha'^3$ correction in large volume limit and genus one amplitude which arises from one loop correction, respectively. This computation is done in $T^6/\mathbb{Z}_N$ orbifold limit of CY$_3$ (see \cite{Antoniadis:1997eg,Antoniadis:2002tr} for similar calculations), where there are some points which remain invariant under discrete $\mathbb{Z}_N$ group of transformations, called orbifold fixed points or EH vertices, where $\chi \neq 0$ \footnote{Although Ref. \cite{Antoniadis:2019rkh} shows that for orbifolds tree level contribution vanishes, which creates a little paradoxical situation.}. They act as the sources of emission of massless gravitons and massive Kaluza-Klein (KK) modes in the internal space. The one loop term in the action of Eq. (\ref{eq:grav_action}) is modified by a correction arising from the effect of exchange of massless as well as massive closed string excitations between EH vertices and $D7$ branes and $O7$ planes, viz., logarithmic correction \cite{Antoniadis:2019rkh}, which takes the form
\begin{equation}\small
\begin{split}
     S_{\mathrm{grav}} &=\frac{1}{(2\pi)^7 \alpha'^4}\int_{\mathcal{M}_4\times\mathcal{X}_6} e^{-2\varphi} R_{(10)}\\
     &+ (\alpha')^3\frac{\chi}{(2\pi \alpha')^4}\int_{\mathcal{M}_4}(2\zeta(3)e^{-2\varphi}\\&+4\zeta (2)(1-\sum_k e^{2\varphi}T_k \ln(R_{\perp}^k /w)))R_{(4)},
\end{split}
\label{eq:mod_grav_action}
\end{equation}
where $T_k$ is the tension of $k$th $D7$ brane, $R_\perp$ stands for size of the transverse 2-cycle volume and $w$ is the width of effective UV cut-off for graviton/KK modes \cite{Antoniadis:2002tr}. Considering all these corrections, the volume term can be re-written as \cite{Haack:2018ufg,Basiouris:2020jgp}
\begin{equation}\small
    \mathcal{V'}= \mathcal{V}+\xi +\sum_{k=1}^3 \eta_k\ln(\tau_k).
\end{equation}
For the sake of simplicity, it is assumed that all branes are identical so that they all have same tension of constant magnitude viz., $e^{-\varphi}T_0$. Therefore,
\begin{equation}\small\small
    \eta_k = \eta = -\frac{1}{2}e^{\varphi}T_0\xi,
    \label{eq:eta_eq}
\end{equation} where $\eta<0$, $\xi,T_0,e^{\varphi} >0$ and $|\eta|<<\xi$ \cite{Antoniadis:2019rkh}.
 Then,
\begin{equation}\small
    \mathcal{V'}=\mathcal{V}+\xi+\eta\sum_{k=1}^3 \ln(\tau_k)=\mathcal{V}+\xi+\eta\ln(\mathcal{V}),
    \label{eq:vol_cor}
\end{equation}
where a factor $2$ is temporarily absorbed in $\eta$. $\xi = -\frac{\chi}{2}\zeta (2) e^{2\varphi}$ for orbifold and $-\frac{\chi}{4}\zeta (3)$ for smooth CY$_3$ \cite{Antoniadis:2019rkh}, that means $\chi<0$ so that $\xi>0$ which will be used later. Now, we can finally write the modified version of Eq. (\ref{eq:tree_kahler}),
\begin{equation}\small
    \mathcal{K} (S,z_a,\rho_k)=-2\ln (\mathcal{V}+\xi+\eta\ln\mathcal{V})-\mathcal{K}_0 (S,z_a)
    \label{eq:mod_kahler_pot}
\end{equation}
where the 6-cycle CY-volume satisfies Eq. (\ref{eq:volume}). As we are interested only in K\"{a}hler moduli stabilization, afterwards we will safely ignore the $\mathcal{K}_0 (S,z_a)$ term (which is also clear from the third term of Eq. (\ref{eq:no_scale_v_F})) and will proceed with only the K\"{a}hler moduli dependent term
\begin{equation}\small
    \mathcal{K}(\rho_k)=-2\ln(\mathcal{V}+\xi+\eta\ln\mathcal{V}).
    \label{eq:use_kh_pot}
\end{equation}
This K\"{a}hler potential breaks the supersymmetric no-scale structure i.e., now,
\begin{equation}\small
    \sum_{k,k' \in  \mathcal{M}^{1,1}({\mathcal{X}_6}) }^3 {\mathcal{K}}^{{\rho_k}\bar{\rho}_{k'}} \partial_{\rho_k} \mathcal{K} \partial_{\bar{\rho}_{k'}} \mathcal{K} \neq 3
\end{equation}
leading to a non-vanishing $F$-term potential
\begin{equation}\small
    V_F = e^{\mathcal{K}}\left(\sum_{k,k' \in \mathcal{M}^{1,1}({\mathcal{X}_6})}^{3} {\mathcal{K}}^{\rho_k \bar{\rho}_{k'}} \mathcal{D}_{\rho_k} \mathcal{W} \mathcal{D}_{\bar{\rho}_{k'}}\bar{\mathcal{W}} - 3\mathcal{W} \bar{\mathcal{W}}\right),
    \label{eq:final_F_pot}
\end{equation}
where $\mathcal{W}$ is the corrected fluxed superpotential of Eq. (\ref{eq:cor_super_pot}). Thus, the K\"{a}hler moduli sector of $\mathcal{X}_6$ is stabilized by perturbative and non-perturbative quantum corrections in classical $F$-term potential. We assume that the non-supersymmetrically stabilized $\tau_{2,3}$ will be just massive enough to be able to sneak out to $\mathcal{M}_4$, for which cosmological inflation will be possible.
Now we split the Eq. (\ref{eq:final_F_pot}) into $V_1$, $V_2$ and $V_3$ such that $V_F = V_1 +V_2 +V_3$:
\begin{equation}\small
     V_1 = e^{\mathcal{K}}\left(\sum_{k,k' \in \mathcal{M}^{1,1}({\mathcal{X}_6})}^{3} {\mathcal{K}}^{\rho_k \bar{\rho}_{k'}} \partial_{\rho_k} \mathcal{K} \partial_{\bar{\rho}_{k'}}\mathcal{K} - 3\right)\mathcal{W} \bar{\mathcal{W}},
\end{equation}
\begin{equation}\small
    V_2 =e^{\mathcal{K}}\sum_{k,k' \in \mathcal{M}^{1,1}({\mathcal{X}_6})}^{3} {\mathcal{K}}^{\rho_k \bar{\rho}_{k'}} \partial_{\rho_k} \mathcal{W} \partial_{\bar{\rho}_{k'}}\bar{\mathcal{W}},
\end{equation}
\begin{equation}\small
\begin{split}
     &V_3 =\\ &e^{\mathcal{K}}\sum_{k,k' \in \mathcal{M}^{1,1}({\mathcal{X}_6})}^{3} {\mathcal{K}}^{\rho_k \bar{\rho}_{k'}}\left(\bar{\mathcal{W}} \partial_{\rho_k} \mathcal{W} \partial_{\bar{\rho}_{k'}}\mathcal{K} + \mathcal{W} \partial_{\rho_k} \mathcal{K} \partial_{\bar{\rho}_{k'}}\bar{\mathcal{W}}\right).
\end{split}
\end{equation}
The term $e^{\mathcal{K}}$ can be approximated using Eq. (\ref{eq:use_kh_pot}) as
\begin{equation}\small
    e^{\mathcal{K}}\approx \frac{1}{\mathcal{V}^2} - \frac{2(\xi +\eta\ln\mathcal{V})}{\mathcal{V}^3} + \frac{6\xi\eta}{\mathcal{V}^4},
    \label{eq:e_term_approx}
\end{equation}
where the terms $\mathcal{O}(\xi^2)$, $\mathcal{O}(\eta^2)$ and $\mathcal{O}(\frac{1}{\mathcal{V}^5})$ are neglected in large volume limit. Similarly, we compute the three terms $V_1$, $V_2$ and $V_3$ using Eqs. (\ref{eq:kh_moduli}), (\ref{eq:volume}), (\ref{eq:cor_super_pot}) and (\ref{eq:use_kh_pot})  in WOLFRAM MATHEMATICA 12 and write the results,
\begin{equation}\small
\begin{split}
     V_1 &= 3e^{\mathcal{K}}(\Tilde{A}+\mathcal{W}_0)^2\\& \frac{(-8\eta+2\xi+2\eta\ln\mathcal{V})\mathcal{V}^2 - 2\eta\mathcal{V}(\eta+2\xi+2\eta\ln\mathcal{V})}{(16\eta-2\xi-2\eta\ln\mathcal{V}+4\mathcal{V})\mathcal{V}^2 +2\eta\mathcal{V}(3\eta+2\xi+2\eta\ln\mathcal{V})},
\end{split}
\end{equation}

\begin{equation}\small
\begin{split}
     V_2 &=4e^{\mathcal{K}}\Tilde{A}^2\alpha^2\tau_1^2\mathcal{V}\\
     &\frac{(2\eta\mathcal{V}(\eta+\xi+\eta\ln\mathcal{V})+2\mathcal{V}^2 (3\eta+\mathcal{V}))(2\eta\ln\mathcal{V}+2(\xi+\mathcal{V}))}{(\mathcal{V}^2+\eta\mathcal{V})(2\eta\mathcal{V}(3\eta+2\xi+2\eta\ln\mathcal{V})+\mathcal{V}^2 (16\eta-2\xi-2\eta\ln\mathcal{V}+4\mathcal{V}))},
\end{split}
\end{equation}
\begin{equation}\small
\begin{split}
     V_3 &= 8e^{\mathcal{K}}\alpha\tau_1 \Tilde{A}(\Tilde{A}+\mathcal{W}_0)\\&\frac{(\mathcal{V}^2 +\eta\mathcal{V})(2\eta\ln\mathcal{V}+2\xi+2\mathcal{V})}{2\eta\mathcal{V}(3\eta+2\xi+2\eta\ln\mathcal{V})+\mathcal{V}^2(16\eta-2\xi-2\eta\ln\mathcal{V}+4\mathcal{V})},
\end{split}
\end{equation}
where $\Tilde{A}=Ae^{-\alpha\tau_1}$. Using Eq. (\ref{eq:e_term_approx}) we binomially expand $V_{1,2,3}$ in large volume limit and release the $2$ factor in $\eta$ (which was absorbed in Eq. (\ref{eq:vol_cor})) to obtain,
\begin{equation}\small
\begin{split}
     V_1 &\approx \frac{3}{2}\mathcal{W}_0^2 \frac{\xi-2\eta (4-\ln\mathcal{V})}{\mathcal{V}^3}-9\mathcal{W}_0^2 \frac{\xi\eta\ln\mathcal{V}}{\mathcal{V}^4}\\& + (2\mathcal{W}_0 \Tilde{A} + \Tilde{A}^2)\left(\frac{3(\xi-2\eta(4-\ln\mathcal{V}))}{2\mathcal{V}^3}-\frac{9\xi\eta\ln\mathcal{V}}{\mathcal{V}^4}\right)\\
     &=V_{1\mathrm{p}} + V_{1\mathrm{m}},
\end{split}
\end{equation}
\begin{equation}\small
\begin{split}
    V_2 &\approx \frac{4\alpha\tau_1\Tilde{A}(\alpha\tau_1\Tilde{A})}{\mathcal{V}^2}\\&-2\alpha\tau_1\Tilde{A}(\alpha\tau_1\Tilde{A})\left(\frac{\xi+2\eta(4+\ln\mathcal{V})}{\mathcal{V}^3}+\frac{2\xi\eta(2-3\ln\mathcal{V})}{\mathcal{V}^4}\right)\\
    &=V_{2\mathrm{np}}+V_{2\mathrm{m}},
\end{split}
\end{equation}
\begin{equation}\small
    \begin{split}
        V_3 &\approx \frac{4\alpha\tau_1\Tilde{A}(\Tilde{A}+\mathcal{W}_0)}{\mathcal{V}^2}-2\alpha\tau_1\Tilde{A}(\Tilde{A}+\mathcal{W}_0)\\&\left(\frac{\xi+2\eta(6+\ln\mathcal{V})}{\mathcal{V}^3}+\frac{6\xi\eta(1-\ln\mathcal{V})}{\mathcal{V}^4}\right)\\
        &=V_{3\mathrm{np}}+V_{3\mathrm{m}},
    \end{split}
\end{equation} where the indices `p', `np' and `m' respectively refer to perturbative, non-perturbative and mixed terms  in $V_{1,2,3}$. Assembling all terms we finally obtain the perturbative, non-perturbative and mixed parts of $V_F$ as,
\begin{equation}\small
    V_{F_1}=V_{1\mathrm{p}}=\frac{3}{2}\mathcal{W}_0^2 \frac{\xi-2\eta (4-\ln\mathcal{V})}{\mathcal{V}^3}-9\mathcal{W}_0^2 \frac{\xi\eta\ln\mathcal{V}}{\mathcal{V}^4},
    \label{eq:old_V_1}
\end{equation}
\begin{equation}\small
    V_{F_2}=V_{2\mathrm{np}}+V_{3\mathrm{np}}=\frac{4\alpha\tau_1}{\mathcal{V}^2}\Tilde{A}(\Tilde{A}+\alpha\tau_1\Tilde{A}+\mathcal{W}_0)
    \label{eq:old_V_2}
\end{equation} and 
\begin{equation}\small
    V_{F_3}=V_{1\mathrm{m}}+V_{2\mathrm{m}}+V_{3\mathrm{m}}=\Tilde{A}(\Tilde{A}f + \mathcal{W}_0 g)
    \label{old_V_3}
\end{equation} where,
\begin{equation}\small
    \begin{split}
        f=&(3\xi-8\eta(2\alpha\tau_1(2\alpha\tau_1+3)+3)-4\xi\alpha\tau_1(\alpha\tau_1+1)\\&-2\eta(2\alpha\tau_1-1)(2\alpha\tau_1+3)\ln\mathcal{V})/(2\mathcal{V}^3)\\
        &+\frac{\eta\xi(2\alpha\tau_1+3)((6\alpha\tau_1-3)\ln\mathcal{V}-4\alpha\tau_1)}{\mathcal{V}^4},
    \end{split}
\end{equation}
 
\begin{equation}\small
\begin{split}
    g=&\frac{(3-2\alpha\tau_1)(\xi+2\eta\ln\mathcal{V})-24\eta(1+\alpha\tau_1)}{\mathcal{V}^3}\\&-6\eta\xi\frac{(3-2\alpha\tau_1)\ln\mathcal{V}+2\alpha\tau_1}{\mathcal{V}^4}.
\end{split}
\end{equation}
To stabilize the smallest K\"{a}hler modulus $\tau_1$ supersymmetrically we first approximate the K\"{a}hler potential as,
\begin{equation}\small
    \mathcal{K}\approx -2\ln\mathcal{V}
\end{equation} by considering the perturbative corrections $\xi$ and $\eta$ to have negligible contributions, which will make calculations simpler. Then equating the covariant derivative of $\mathcal{W}$ of Eq. (\ref{eq:cor_super_pot}) w.r.t $\rho_1$ to zero at $\rho_1 = i\tau_1$ we get,
\begin{equation}\small
\begin{split}
     &D_{\rho_1}\mathcal{W}|_{\rho_1=i\tau_1}=ie^{-\alpha\tau_1}\left(\alpha A+\frac{A+\mathcal{W}_0 e^{\alpha\tau_1}}{2\tau_1}\right)=0\\
     &\mathrm{or,}\quad (-\alpha\tau_1-\frac{1}{2})e^{(-\alpha\tau_1-\frac{1}{2})}=\frac{\mathcal{W}_0}{2A\sqrt{e}}\in \mathbb{R}.
     \label{eq:stabl_tau}
\end{split}
\end{equation}
The physical solution of this equation is called the `Lambert W-function', $w=W_{0}(\frac{\mathcal{W}_0}{2A\sqrt{e}})$ corresponding to the $0$-branch \cite{Basiouris:2020jgp}.
Therefore,
\begin{equation}\small
 -\alpha\tau_1-\frac{1}{2}=w\\
    \longrightarrow\tau_1=-\frac{1+2w}{2\alpha}.
\label{eq:tau}
\end{equation} Here $\tau_1$ will be chosen as 40 as in \cite{Basiouris:2020jgp}. Also from Eq. (\ref{eq:stabl_tau}) we get,
\begin{equation}\small
    \Tilde{A}=Ae^{-\alpha\tau_1}=-\frac{\mathcal{W}_0}{1+2\alpha\tau_1}=\frac{\mathcal{W}_0}{2w}.
    \label{eq:A_tilde}
\end{equation}
Let, $\epsilon = \frac{1+2w}{w}\quad (\approx 1$) \cite{Basiouris:2020jgp}, then from Eqs. (\ref{eq:tau}) and (\ref{eq:A_tilde}) we obtain,
\begin{equation}\small
    2\alpha\tau_1=-\frac{\epsilon}{\epsilon-2}
    \label{eq:mod_two_alpha}
\end{equation} and
\begin{equation}\small
    \Tilde{A}=\frac{\epsilon-2}{2}\mathcal{W}_0.
    \label{eq:mod_a_tilde}
\end{equation}
Using Eqs. (\ref{eq:mod_two_alpha}) and (\ref{eq:mod_a_tilde}) we can rewrite Eqs. (\ref{eq:old_V_1}), (\ref{eq:old_V_2}) and (\ref{old_V_3}) as,
\begin{equation}\small
    V_{F_1}=\frac{3}{2}\mathcal{W}_0^2\frac{\xi-2\eta(4-\ln\mathcal{V})}{\mathcal{V}^3}-9\mathcal{W}_0^2\frac{\xi\eta\ln\mathcal{V}}{\mathcal{V}^4},
\end{equation}
\begin{equation}\small
    V_{F_2}=-(\epsilon\mathcal{W}_0)^2\frac{\mathcal{V}}{4\mathcal{V}^3}
\end{equation} and 
\begin{equation}\small
\begin{split}
      V_{F_3}&=(\epsilon\mathcal{W}_0)^2\left(\frac{2\xi+4\eta(\ln\mathcal{V}-1)}{4\mathcal{V}^3}-\eta\xi\frac{3\ln\mathcal{V}-1}{\mathcal{V}^4}\right)\\&-\left(\frac{3}{2}\mathcal{W}_0^2\frac{\xi-2\eta(4-\ln\mathcal{V})}{\mathcal{V}^3}-9\mathcal{W}_0^2\frac{\xi\eta\ln\mathcal{V}}{\mathcal{V}^4}\right).
\end{split}
\end{equation}
Now we can finally write the $F$-term potential as,
\begin{equation}\small
\begin{split}
    V_F &= V_{F_1}+V_{F_2}+V_{F_3}\\&=-(\epsilon\mathcal{W}_0)^2\left(\frac{\mathcal{V}-2\xi+4\eta(1-\ln\mathcal{V})}{4\mathcal{V}^3}-\eta\xi\frac{1-3\ln\mathcal{V}}{\mathcal{V}^4}\right)
\end{split}
    \label{eq:Final_F_term_pot}
\end{equation} which matches with the corresponding expression in \cite{Basiouris:2020jgp}.
This $F$-term potential arises because of deviation from the no-scale structure of the tree level K\"{a}hler potential vis-\`{a}-vis the supersymmetry breaking through non-perturbative quantum correction of fluxed superpotential by the smallest K\"{a}hler modulus $\tau_1$, which ensures that $\tau_1$ is stabilized supersymmetrically and lies in compactified dimensions. The other two moduli $\tau_{2,3}$ are perturbatively stabilized through overall volume correction of the internal space $\mathcal{X}_6$. Being just massive enough, they can come out of the compactified dimensions and appear in non-compact 4-manifold $\mathcal{M}_4$. These relatively larger moduli $\tau_{2,3}$ play very significant role in cosmology. Certain symmetric combination of them will give rise to the inflaton field $\phi$ which drives the inflation by providing required large vacuum energy, whereas their anti-symmetric combination is another field which will remain  in the background. Now, it is quite clear from Eq. (\ref{eq:Final_F_term_pot}) that this $V_F$ can not alone be responsible for cosmological inflation because its minima is at $AdS$ space. It requires some uplifting agent to obtain a slow-roll like $dS$ vacuum.
There are actually so many uplifting mechanisms which include applying $\bar{D3}$ brane \cite{Baumann:2010sx}, Fayet-Iliopoulos (FI) $D$-term \cite{Cribiori:2017laj} and nilpotent superfield \cite{Ferrara:2014kva}. Here we use $D$-term contributions associated with $U(1)$ factors in the models of intersecting $D7$ branes \cite{Antoniadis:2019doc,Antoniadis:2018hqy,Cremades:2007ig,Haack:2006cy,Burgess:2003ic} and the corresponding $D$-term potential takes the form \cite{Antoniadis:2018hqy,Basiouris:2020jgp}
\begin{equation}\small
    V_D=\sum_{i=1}^3\frac{g_{i}^2}{2}\left(\sqrt{-1}Q_i\partial_{\rho_i}\mathcal{K}+\sum_{j}q_j|\langle\Phi_j\rangle|^2\right)^2 \approx \sum_{i=1}^3 \frac{d_i}{\tau_i^3},
    \label{eq:D_term_pot}
\end{equation}
where $g_i$'s (${g_i}^{-2}=\tau_i$ + flux dependent and curvature corrections involving dilaton \cite{Haack:2006cy}) are the $U(1)$ gauge couplings and $d_i={Q_i}^2/8>0$ ($i=1,2,3$) correspond to the charges carried out by the 7-branes. $q_j$'s are the charges of the matter fields $\Phi_j$'s whose VEVs are considered to be zero. This is a valid approximation because we are considering only the gravitational sector (see Eq. (\ref{eq:grav_action})) and also this is a safe and simple assumption for obtaining a non-vanishing $V_D$ (see \cite{Haack:2006cy} and \cite{Burgess:2003ic} for more details).
$\frac{d_1}{\tau_1^3}$ acts as constant uplifting factor similar to FI $D$-term. Remaining two terms will manifest through inflaton and an auxiliary field. The effective potential can now be expressed using Eqs. (\ref{eq:Final_F_term_pot}) and (\ref{eq:D_term_pot}) as
\begin{equation}\small
\begin{split}
    V_{\mathrm{eff}}&=V_F + V_D=\\&-(\epsilon\mathcal{W}_0)^2\left(\frac{\mathcal{V}-2\xi+4\eta(1-\ln\mathcal{V})}{4\mathcal{V}^3}-\eta\xi\frac{1-3\ln\mathcal{V}}{\mathcal{V}^4}\right)\\&+ \sum_{i=1}^3 \frac{d_i}{\tau_i^3}.
\end{split}
    \label{eq:final_eff_pot}
\end{equation}
Let us transform $\tau_{2,3}$ into two canonically normalized fields $t_{2,3}$ as
\begin{equation}\small
        t_2 =\frac{1}{\sqrt{2}}\ln(\sqrt{\tau_1}\tau_2),\quad t_3=\frac{1}{\sqrt{2}}\ln(\sqrt{\tau_1}\tau_3)
\end{equation}
where $\tau_1$ is considered to be constant according to Eq. (\ref{eq:tau}) as it is supersymmetrically stabilized. A symmetric combination of $t_{2,3}$ yields
\begin{equation}\small
        \phi = \frac{1}{\sqrt{2}}(t_2+t_3)
        =\frac{1}{2}\ln(\tau_1\tau_2\tau_3)
        =\ln\mathcal{V}.
        \label{eq:first_inf}
\end{equation}
Thus, the inflaton field manifests as logarithm of the CY$_3$ volume. Also from Eq. (\ref{eq:first_inf}) we can write
\begin{equation}\small
    \tau_2\tau_3 = \frac{e^{2\phi}}{\tau_1}.
    \label{eq:multi_tau_23}
\end{equation}
By antisymmetrizing $t_{2,3}$ we obtain
\begin{equation}\small
   u=\frac{1}{\sqrt{2}}(t_2-t_3)=\frac{1}{2}\ln\left(\frac{\tau_2}{\tau_3}\right)\longrightarrow\quad \frac{\tau_2}{\tau_3}=e^{2u}.
 \label{eq:div_tau_23}
\end{equation}
These $\phi$ and $u$ fields are just two new avatars of $\tau_{2,3}$ and therefore we can express $\phi$ and $u$ in terms of $\tau_{2,3}$ by inverting Eqs. (\ref{eq:multi_tau_23}) and (\ref{eq:div_tau_23}) as
\begin{equation}\small
    \tau_2=\frac{e^{(\phi+u)}}{\sqrt{\tau_1}},\quad \tau_3=\frac{e^{(\phi-u)}}{\sqrt\tau_1}.
    \end{equation}
Now, we can transform the effective potential of Eq. (\ref{eq:final_eff_pot}) as a 2-field potential (see Figure \ref{fig:3D})
\begin{equation}\small
\begin{split}
     &V_{\mathrm{eff}}(\phi,u)=-(\epsilon\mathcal{W}_0)^2\left(\frac{e^{\phi}-2\xi+4\eta(1-\phi)}{4e^{3\phi}}-\eta\xi\frac{1-3\phi}{e^{4\phi}}\right)\\&+\frac{d_1}{\tau_1^3}+\tau_1^{3/2}(d_2e^{-3(\phi+u)}+d_3e^{-3(\phi-u)}).
\end{split}
\label{eq:2_field_inf_pot}
\end{equation} Now, in order to stabilize the $u$ field we set,
\begin{equation}\small
    \left(\frac{\partial V_{\mathrm{eff}}(\phi,u)}{\partial u}\right)_{u=u_0}=0 \longrightarrow u_0 = \frac{1}{6}\ln\left(\frac{d_2}{d_3}\right)
   \end{equation} and 
\begin{equation}\small
\left(\frac{\partial^2V_{\mathrm{eff}}(\phi,u)}{\partial u^2}\right)_{u=u_0}=9d\tau_1^{3/2}e^{-3\phi}>0
\end{equation} where $d=2\sqrt{d_2d_3}$. Now, we finally obtain the single field slow-roll inflaton potential with a stable $dS$ vacuum,
\begin{equation}\small
\begin{split}
    &V(\phi,u_0)\equiv V(\phi)=\eta(\epsilon\mathcal{W}_0)^2e^{-3\phi}\\&\left[\phi+\left(\frac{\xi}{2\eta}-1+\frac{d\tau_1^{3/2}}{\eta\epsilon^2\mathcal{W}_0^2}\right)-\frac{e^\phi}{4\eta}+\xi e^{-\phi}(1-3\phi)\right]+\frac{d_1}{\tau_1^3}.
\end{split}
    \label{eq:final_ds_inf_pot}
\end{equation} We can compress this equation by considering
\begin{equation}\small
    \alpha=\eta(\epsilon\mathcal{W}_0)^2,\beta=\left(\frac{\xi}{2\eta}-1+\frac{d\tau_1^{3/2}}{\eta\epsilon^2\mathcal{W}_0^2}\right),\\
        \gamma=\frac{1}{4\eta},\lambda=\frac{d_1}{\tau_1^3}
    \label{eq:string_param}
\end{equation} as
\begin{equation}\small
\boxed{V(\phi)=\alpha e^{-3\phi}\left[\phi+\beta-\gamma e^{\phi}+\xi e^{-\phi}(1-3\phi)\right]+\lambda}.
\label{eq:compressed_final_ds_inf_pot}
\end{equation}
Our derived inflaton potential $V(\phi)$ (Eq. (\ref{eq:compressed_final_ds_inf_pot})) crucially depends on four parameters $\alpha$, $\beta$, $\gamma$ and $\lambda$ which in turn depend on perturbative and non-perturbative string theoretic parameters: $\epsilon$, $\mathcal{W}_0$, $\xi$, $\eta$, $g_s$, $T_0$, $\chi$, $\tau_1$, $d_1$ and $d$ according to the Eq. (\ref{eq:string_param}). In our framework we choose these parameters as follows:
\begin{table}[H]\small
    \centering
    \begin{tabular}{|c|c|c|c|c|c|c|c|}
    \hline
         $\epsilon$ & $\mathcal{W}_0$ & $\xi$ & $\eta$ & $g_s$ & $T_0$ & $\chi$ & $\tau_1$ \\
         \hline
         \hline
         1 & 1.59 & 23 & -0.71 & 0.6 & 0.103 & -77 & 40  \\
         1 & 1.59 & 52 & -0.71 & 0.6 & 0.045 & -173 & 40 \\
         1 & 1.59 & 65 & -0.71 & 0.6 & 0.037 & -217 & 40 \\
         1 & 1.59 & 65 & -0.71 & 0.6 & 0.037 & -217 & 40 \\
         1 & 1.59 & 65 & -0.71 & 0.6 & 0.037 & -217 & 40 \\
         1 & 1.59 & 80 & -0.71 & 0.6 & 0.029 & -267 & 40 \\
         \hline
    \end{tabular}
    \caption{In our first parameter space we have chosen $\epsilon$, $\mathcal{W}_0$, $\eta$, $g_s$ and $\tau_1$ to be fixed parameters while $\xi$, $T_0$ and $\chi$ are varied. The parameters are consistent with the constraints given in \cite{Basiouris:2020jgp,Antoniadis:2019rkh}. $T_0$ and $\chi$ are parameterized in such a way so that $\eta$ remains almost fixed. }
    \label{tab:tab_params_1}
\end{table}
\begin{table}[H]\small
    \centering
    \begin{tabular}{|c|c|c|c|c|c|}
    \hline
          $\alpha$ & $\beta$ & $\gamma$ & $\lambda$ & $d_1$ & $d$  \\
          \hline
          \hline
          -1.805 & -70 & -0.35 & $5.47\times 10^{-6}$ & 0.35 & 0.45\\
           -1.805 & -80 & -0.35 & $5.47\times 10^{-6}$ & 0.35 & 0.45\\
            -1.805 & -90 & -0.35 & $5.47\times 10^{-6}$ & 0.35 & 0.17\\
             -1.805 & -100 & -0.35 & $5.47\times 10^{-6}$ & 0.35 & 0.31\\
              -1.805 & -110 & -0.35 & $5.47\times 10^{-6}$ & 0.35 & 0.45\\
               -1.805 & -120 & -0.35 & $5.47\times 10^{-6}$ & 0.35 & 0.45\\
          \hline
    \end{tabular}
    \caption{In our second parameter space we have treated $\alpha$, $\gamma$, $\lambda$ and $d_1$ as constants and $\beta$, $d$ to be variables. The $\alpha$, $\beta$, $\gamma$ and $\lambda$ are obtained from the Eq. (\ref{eq:string_param}) using the parameters in Table \ref{tab:tab_params_1} and $d_1$, $d$ are suitably fixed to yield the slow-roll structure of the potential. Although $d_1$ satisfies the condition given in \cite{Basiouris:2020jgp}.}
    \label{tab:tab_params_2}
\end{table} In Figure \ref{fig:ads_potential} we have shown the inflaton potential $V(\phi)$ against the inflaton field $\phi$ from Eq. (\ref{eq:compressed_final_ds_inf_pot}) without the uplifting term $\lambda$. This potential has a plateau-type slow-roll feature with an $AdS$ minimum at $\phi\approx 6.02$, which can not drive the inflationary expansion. In Figures \ref{fig:dspot1}  we have described the actual inflaton potentials in $dS$ space for two sets of the parameter $\beta$ (see Eq. (\ref{eq:string_param})): one is $\beta=-70, -90, -110$ for three values of the non-perturbative parameter $d=0.17, 0.31, 0.45$ respectively keeping the perturbative parameter fixed at $\xi=65$ (see upper figure) and the other is $\beta=-80,-100, -120$ by varying $\xi=23, 52, 80$ respectively for a particular value of $d=0.45$ (see lower figure). Both the figures highlight an uplifting and a slight shift in $\phi$ direction of the $dS$ vacuum by the increase of $\beta$ maintaining the same slow-roll plateau as found in $AdS$ space. We find that, in these two figures, uplifting the $dS$ vacuum does not disturb the flat direction, necessary for inflation. It is observed that the two perturbative parameters $\xi$ and $\eta$ play a major role for shaping the inflaton potential as slow-roll one, the non-perturbative parameters $d$ and $d_1$ are responsible for the uplifting and the $\alpha$ fixes the overall energy scale of inflation which is $\sim 10^{-6}$ in our case. The smallness of this energy scale firmly indicates the microscopic origin of our inflaton potential viz., the moduli stabilization in type IIB/F-theory compactification with string, brane, orientifold and fluxes- which is certainly a prime motivation of our approach.  
\begin{figure}[H]\small
    \centering
    \onefigure[width=0.5\linewidth]{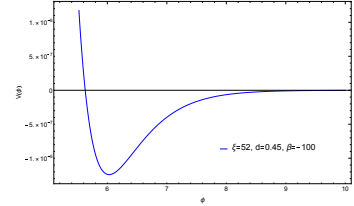}
    \caption{Inflaton potential with $AdS$ minima at $\phi\approx 6.02$ for $\xi=52$, $d=0.45$ and $\beta=-100$.}
    \label{fig:ads_potential}
\end{figure}
\begin{figure}[H]\small
  \centering
    \onefigure[width=0.5\linewidth]{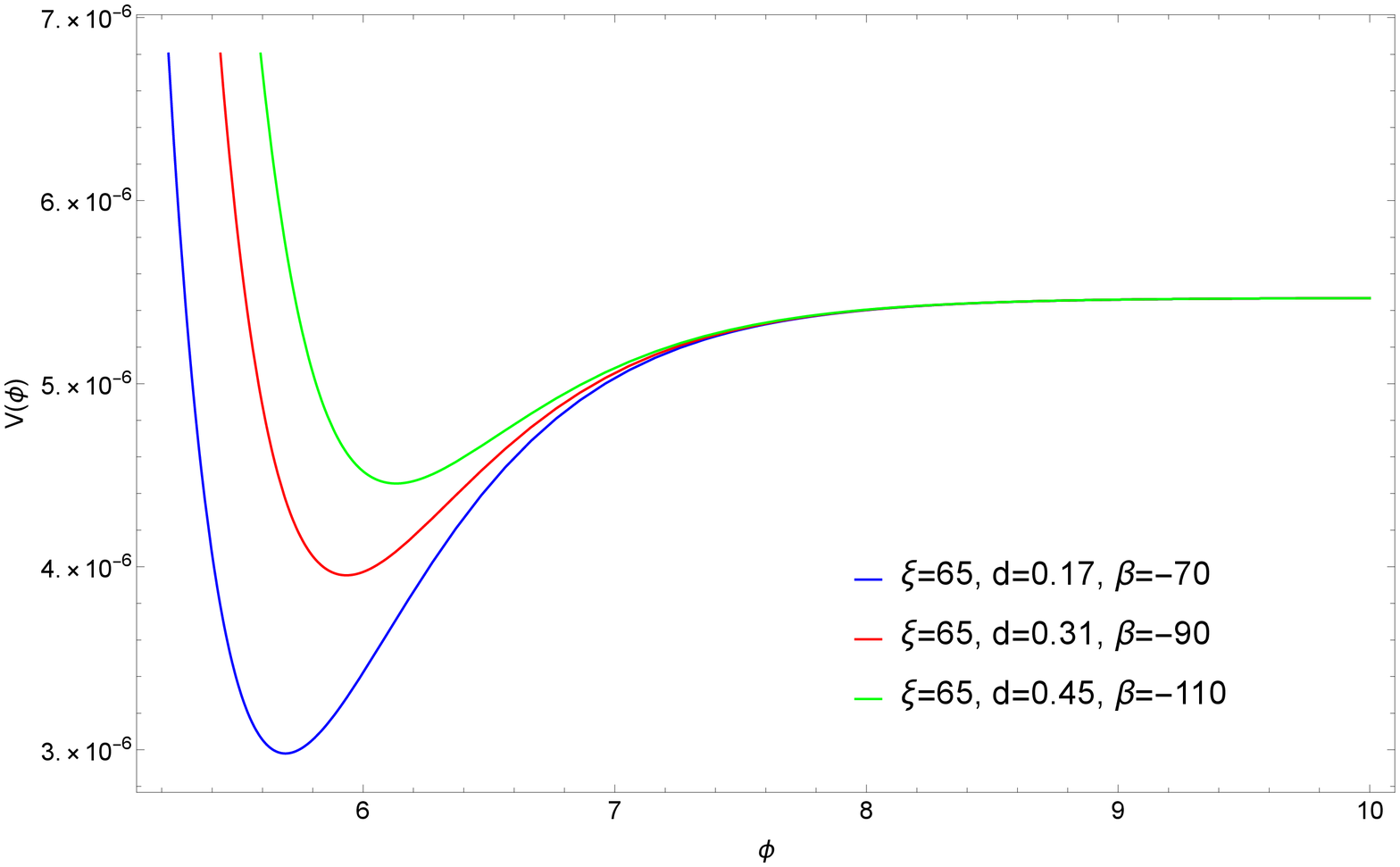}
     \onefigure[width=0.5\linewidth]{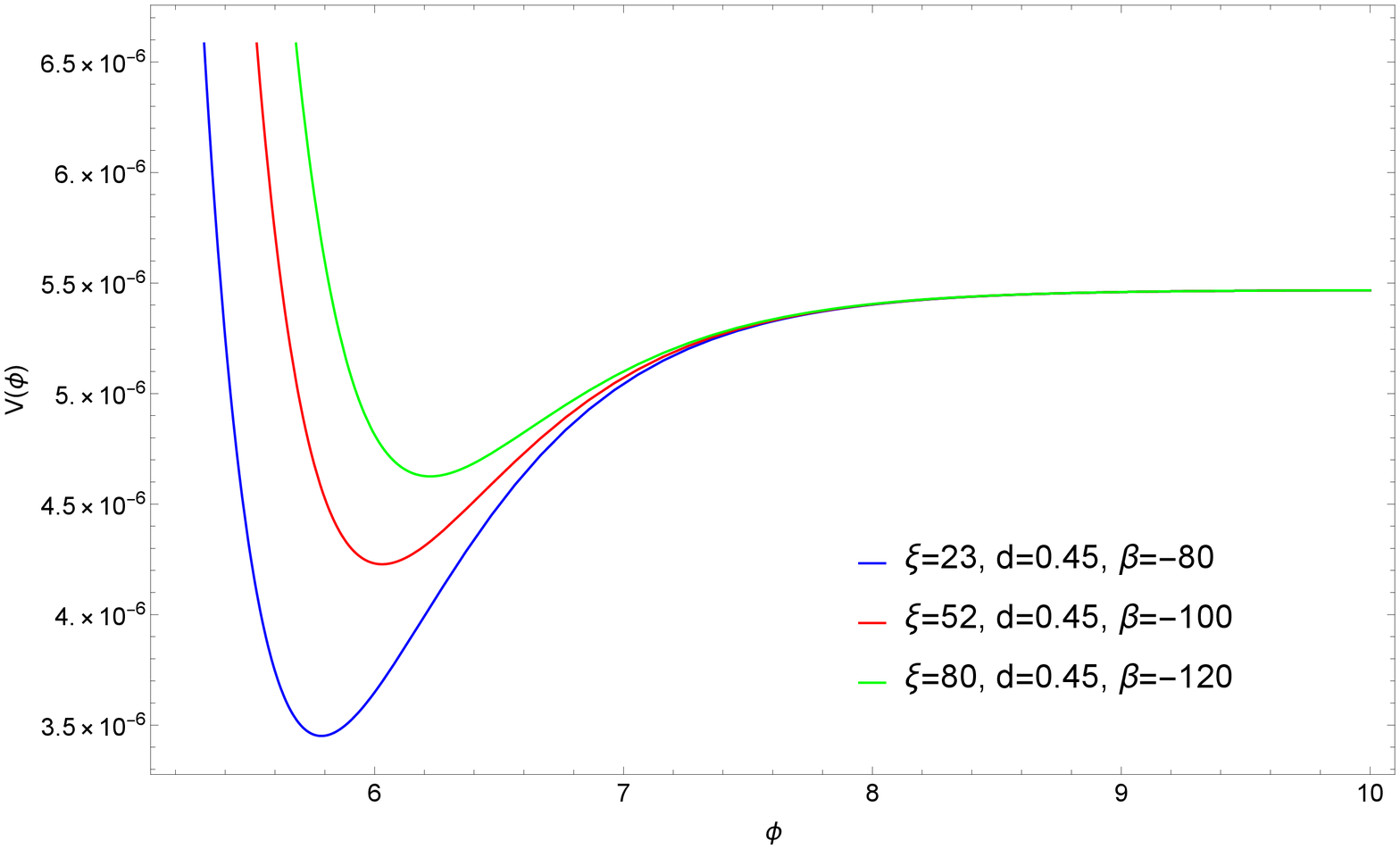}
    \caption{Uplifted inflaton potentials with $dS$ minima for different values of $\beta$, $d$ with fixed value of $\xi$ (upper figure) and different values of $\beta$, $\xi$ with fixed value of $d$ (lower figure). The $dS$ vacua are uplifted as well as shifted in right direction keeping the slow-roll plateau same.}
    \label{fig:dspot1}   
   \end{figure}
 At the end, we would like to mention that with the slow-roll potential of Eq. (\ref{eq:compressed_final_ds_inf_pot}) and following the formalism in Ref. \cite{Sarkar:2021ird} we have obtained the values of some cosmological parameters such as, scalar power spectrum ($\Delta_s$): $3.38\times 10^{-4}$ - $3.60\times 10^{-4}$, tensor power spectrum ($\Delta_t$): $2.1015\times 10^{-7}$ - $2.1018\times 10^{-7}$, number of e-folds ($N$): $55.0$ - $56.7$, scalar spectral index ($n_s$): $0.9652$ - $0.9662$, tensor spectral index ($n_t$): $(-7.28\times 10^{-5})$ - $(-7.76\times 10^{-5})$ and tensor-to-scalar ratio ($r$): $5.8\times 10^{-4}$ - $6.2\times 10^{-4}$ at $k=0.001$ - $0.009$ Mpc$^{-1}$ for $\xi=52$, $d=0.45$ and $\beta=-100$. We plan to report the details of these calculations in a future publication.
\par In conclusion, we have derived, effectively, a single-field slow-roll inflaton potential from K\"{a}hler moduli stabilizations in type IIB/F-theory.

\acknowledgments
The authors acknowledge the University Grants Commission for the CAS-II program. AL acknowledges CSIR, the Government of India for NET fellowship. AS and CS acknowledge Government of West Bengal for granting them Swami Vivekananda fellowship.

\bibliographystyle{eplbib}
\bibliography{biblio}

\end{document}